\documentclass[aps,prd,showpacs,twocolumn]{revtex4}
\usepackage{graphicx}
\usepackage{dcolumn}
\usepackage{bm}

\begin{document}

\title{Meson2006 Summary: Theory}

\author{T.Barnes}

\affiliation{
Physics Division, Oak Ridge National Laboratory,
Oak Ridge, TN 37831, USA\\
Department of Physics and Astronomy, University of Tennessee,
Knoxville, TN 37996,
USA.}

\begin{abstract}
This is a summary of theoretical plenary contributions to the
biennial hadron physics conference Meson2006, which was the
ninth in this series. The topics covered in the meeting include
low energy pion-pion and pion-nucleon interactions, photoproduction
and hadronic production of light mesons and baryons,
in-medium effects, recent developments in
charmed mesons, charmonia and B mesons, the status of exotica,
and some related topics such as final state interactions. In this contribution
we review and summarize the plenary talks presented by theorists
at the meeting, and emphasize some of the main points of their presentations.
Where appropriate we will add brief comments on some aspects of QCD spectroscopy.
Finally, following tradition, we conclude with a new Feynman story.

\end{abstract}


\maketitle

\section{Introduction}	

The biennial Meson conference series has historically been hosted by a long-standing collaboration
between the German and Polish nuclear and hadron physics communities, especially involving
research institutions at Krak\'ow and J\"ulich. The range of topics represented
at this meeting since the inception in 1991 has grown to span much of the physics
of QCD spectroscopy. As a result of the history of this meeting, there was considerable coverage of
the physics of mesons below 1 GeV, strange mesons, light $u,d,s$ baryon spectroscopy, and
in-medium effects. The recent activity in other areas of QCD, such as the QGP searches at RHIC
and the new discoveries in charm spectroscopy, have led to new research areas in hadron physics
that are now also major components of this meeting.

Here we summarize the plenary theory contributions at Meson2006, as well as some
especially theory-relevant experimental presentations, ordered according to the
broad topics discussed above: light hadrons, medium effects, heavy quark
hadron spectroscopy and multiquark physics.
Of course there is considerable overlap between these areas in some
contributions, which I will stress where appropriate.

\subsection{Plenary contributions on light hadrons}

\subsubsection{Prof.~Leutwyler and the Red Dragon}

We begin our summary with the presentation on the lightest of hadrons.
H.Leutwyler \cite{Leu,Caprini:2005zr}
discussed the status of the $\sigma$ meson (the ``{\it Rote Drache}") in $\pi\pi$ scattering.
Leutwyler noted that the Roy equations,
combined with dispersion relations, give an accurate numerical description of low-energy $\pi\pi$
scattering in terms of just three numbers, which he takes to be the scattering lengths $a_0$,
$a_2$ and a phase shift $\delta$. In this talk Leutwyler reported the interesting new result that
the location of scattering amplitude poles on the 2$^{nd}$ sheet is accurately determined by the
(more accessible) location of zeros on the 1$^{st}$ sheet. This new approach allows a much more accurate
determination of the location of the $\sigma$ pole, with the result
\begin{equation}
{\rm Pole\ position} = 441(4) - 272(6)i \ {\rm MeV}.
\end{equation}
Since there are twin poles in this approach with equal magnitude $\pm$ imaginary parts, 
and they are at the low-energy ``head"
of the I=0 $\pi\pi$ scattering amplitude in the complex E plane, it is amusing to refer to these
poles as the eyes of the red dragon. Thus, what Prof.~Leutwyler has discovered
can be summarized in Latin, as appropriate for a conference in a famous medieval city, as the
\begin{center}
``{\it Loci Oculorum Draconis Rutili} ".
\end{center}
Although the location of the eyes of the dragon is now well established, the actual nature of the
associated $\sigma$ meson state is unfortunately not specified by this method. Thus we now know
{\it where} the $\sigma$ is, but not {\it what} it is.

\subsubsection{Kd scattering}

Since the $K^-p$ and $K^-n$ scattering amplitudes near threshold
are at least moderately well-known experimentally, one may use this
information to evaluate the $K^-d$ scattering length. This theoretical study was
reported by A.Gal \cite{Gal,Gal:2006cw}, and is an improvement on previous work in not using a
fixed-center approximation. Their initial conclusion is that the (complex) $K^-d$
scattering length is fairly large, $\sim 1-2$~fm. Based on this work, they quote an ultimate
goal for the accuracy of this scattering length of $\approx 10\%$ for theory, and
(at the DEAR/SIDDHARTA experiment at DA$\Phi$NE) $\approx 5\%$ for experiment.

\subsubsection{Isospin violation in the strong interaction}

In this contribution, Niskanen discussed prospects for observing evidence for
isospin violation in the strong interaction in the processes
$NN\to d\pi$ \cite{Nis,Machner:2005ha}.
(We use
$N$ generically to represent a nucleon, and $p$ and $n$ specifically for protons and
neutrons.) Of course relatively weak isospin violation is expected in the strong
interaction from several sources, such as electromagnetic corrections, the $u,d$ quark
mass differences, and (presumably indirectly due to these effects) the $p,n$
and other hadron isomultiplet mass differences. In the naive isospin limit one has
\begin{equation}
\frac{pp\to d\pi^+}{pn\to d\pi^0} = 2.
\end{equation}
Although one can search for violations of this ratio, there are various
complications such as the choice of the kinematic point at which the cross sections
are to be compared, and the difference in quality of $p$ and $n$ beams. After a review
of these complications and a long and lively discussion with the audience, Niskanen
offered a new uncertainty principle, which relates the ease $E$ of the experiment and
the associated theoretical interpretation:
\begin{equation}
E_{expt.} \cdot E_{theor.} \geq \ constant.
\end{equation}
\subsubsection{Baryon resonance photoproduction, $\gamma N \to \eta N$ and $\eta' N$.}

Although this was an experimental contribution, there were sufficiently interesting
theoretical issues discussed to merit a mention in this theory summary. In this talk
Tiator \cite{Tia} discussed the photoproduction of $\eta N$ and $\eta' N$ final states, using the
Mainz isobar analysis program \cite{Tiator:2006dq} for pseudoscalar photoproduction.
This reaction was assumed to be dominated by two Feynman diagrams, $t$-channel vector meson exchange
($\rho$ and $\omega$ are included) and $s$-channel nucleon and $N^*$ resonance
production. Two interesting and rather unsettling observations made by Tiator were that

\vskip 0.5cm
\noindent
1) the $D_{15}(1675) \to \eta N$ branching fraction in the fit varies from
$0.7\%$ to $17\%$, depending on the assumed $t$-dependence of vector exchange (Regge versus
form factors), and

\vskip 0.5cm
\noindent
2) the best fit gives a tiny $NN\eta$ coupling,
with $g_{NN\eta}^2 / 4\pi$ less that an order of magnitude smaller than would be expected
from SU(3) symmetry.

\vskip 0.5cm
\noindent
Since the quark model (see Table~I of Downum {\it et al.} \cite{Downum:2006re})
and the Nijmegen $NN$ force model \cite{Nagels:1978sc}
both anticipate that $g_{NN\eta}$
should be comparable to $g_{NN\pi}$, which involves a simple
SU(3) flavor factor, this suggests that the experimental analysis may not be giving realistic
couplings. Whatever the explanation, this is certainly a striking result.

\subsubsection{Baryon resonance photoproduction, $\gamma N \to \phi\eta N$.}

Soyeur \cite{Soy,Soyeur:2006dd} discussed a very interesting baryon resonance photoproduction process,
$\gamma N \to \phi\eta N$. In her theoretical analysis she assumed that this reaction
at low energies is dominated by $t$-channel $\pi$ and $\eta$ meson exchange, with the
$\phi$ meson produced through $\gamma \pi \phi$ and $\gamma \eta \phi$ vertices.
(These of course have known couplings.) The final $\eta$ is assumed to be produced
by an intermediate $N^*(1535)$. This is a nice example of a coupled-channel process,
since it is sensitive to the interference of the off-diagonal amplitude for
$\pi N \to N^*(1535) \to \eta N$
and the diagonal amplitude for
$\eta N \to N^*(1535) \to \eta N$. An experimental study of this reaction can
also establish the relative phase of the $\pi N$ and $\eta N$ couplings of the
$N^*(1535)$, which is predicted by the quark model. Soyeur suggests that this would be
an interesting reaction to study at JLAB, with E$_{\gamma}\approx 4$~GeV.

\subsection{Plenary contributions on in-medium effects}

There were several experimental contributions on in-medium effects on hadron masses and
widths, especially on vector mesons, by Weygand, Pietraszko, Yokkaichi and Krusche.
Although this is certainly an interesting and active topic, one had an uncomfortable initial
feeling during these presentations that the theoretical underpinning of this field was not strong.
Each experimental talk made reference to the same iconic figure of the $\langle \bar q q\rangle$ condensate
versus temperature and density, and stated that this somehow explained the fall in meson mass
with increasing density, without a critical discussion of the physics issues. Even if this picture is
correct, experiments do not measure a condensate; they measure a cross section, a
resonance mass and width, or some other set of observables. The interpretation of
mass shifts in terms of a $\langle \bar q q\rangle$ condensate may be misleading or even incorrect,
and is certainly {\it not} what experiments on in-medium hadron properties themselves contribute.

In view of this rather uncritical experimental support for a theoretical concept, the subsequent
theoretical review of in-medium effects by Mosel \cite{Mos} was refreshing indeed. Mosel noted
that the in-medium hadron mass shifts are not magic, instead they are (mostly)
simple FSI effects \cite{Gallmeister:2006yt}.
Mosel {\it et al.} have developed codes to simulate these FSI effects, and
find that several of the well known medium modifications can easily be explained in this manner.
One example is the well-established doubling of the $\Delta$ width, which they find is
due to $\Delta N$ final state interactions
($\Delta N\to NN$ and $\Delta NN\to NNN $). Another is the downwards
mass shift of the $\pi\pi$ system; ``A big part of the $\pi\pi$ mass shift is due to
$\pi N$ rescattering on the way out." Finally, Mosel warns that the relation between
these mass shifts and $\langle \bar q q\rangle$ is obscure at best: ``The connection
of any hadron mass with the fall-off of $\langle \bar q q\rangle$ is very indirect."

\subsection{Plenary contributions on heavy-quark hadrons}

The very exciting recent developments in the spectroscopy of charm were also discussed
at Meson2006. As these were almost exclusively experimental talks, they are not reviewed
here, and will instead be summarized in the experimental review by Seth \cite{Set}.
The specific topics covered were the
X(1835) (C.Zhang; this was actually a presentation on a light hadrons topic),
B decays to light mesons (I. Gough Eschrich),
$c$ mesons at Belle (S.Korpar)
charmonium at HERA (A.Meyer),
$c$ and $c\bar c$ spectroscopy at BABAR (M.Peliz\"aus),
D$_s$ and  $c\bar c$ spectroscopy at CLEO (M.Shepherd)
and $c$ physics at FOCUS (C.G\"obel).

The single theoretical talk in this area was by A.Krassnigg \cite{Kra},
who reviewed developments in a Dyson-Schwinger approach to meson spectroscopy \cite{Holl:2005vu}.
In this approach one solves coupled integral
equations for self-energy and vertex functions, given a
truncation scheme for $n$-point functions.
Their approach has the advantage that it respects chiral symmetry and is
relativistic. Thus far it has been applied to ground-state pseudoscalars
and vectors, and is now being extended to radial excitations, scalars
and axial vectors. This is a ``work in progress", for example their current
result for the $\eta_c'$ mass is a rather low 3.45~GeV.

\subsection{Plenary contribution on multiquarks}
As a final physics topic,
the experimental presentation by Kabana \cite{Kab,Kabana:2005tq}
on the status of pentaquarks raised some familiar
questions about multiquark states, which may be worth stressing yet again
in this summary. As was noted in the HADRON05 Theory Summary \cite{Barnes:2005zy},
there is a fundamental theoretical problem with light multiquark resonances such as the pentaquark,
which despite three decades of work in this area is still not widely appreciated.
One is reminded of a quote from a Mel Brooks movie regarding earlier reports
of monstrous scientific discoveries;

\vskip 3mm

\noindent
\hskip5mm
``These are very serious charges you're making,

\noindent
\hskip5mm
and all the more painful to us, your elders, because

\noindent
\hskip5mm
we still have nightmares from five times before."

\vskip 2mm
\noindent
\hskip2cm
- village elder, {\it Young Frankenstein} \cite{Brooks}
\vskip 3mm

The problem with multiquarks is that if a multiquark system such as
$q^2\bar q^2$ is above the threshold for dissociation into two $q\bar q$
mesons (such as a J$^{PC} = 0^{++}$ $u^2\bar d^2$ above $2m_{\pi}$,
or a $\theta(1540)$ $q^4\bar s$ pentaquark above $KN$),
this system can simply ``fall-apart" into two lighter hadrons without a decay
interaction.
Although this has entered physics folklore as the notion that
multiquark systems will be very broad, the actual effect of fall-apart
modes below threshold may be much more drastic; without a fission barrier
to suppress dissociation, these multiquark systems need not form resonances at all.
This option, the absence of light multiquark resonances due to fall-apart modes,
was appreciated in some of the theoretical literature on multiquarks in the
late 1970s \cite{Jaffe:1976yi},
but is in danger of being forgotten. For this reason, Isgur's warning about the
recurring ``Multiquark Fiasco" \cite{Isgur:1985vp}
should be required reading for anyone working on the physics of multiquarks.

\section{Feynman Story}

This section continues my ``tradition" of ending Meson with a new Feynman story 
\cite{Barnes:2000rr}.
The story of the pentaquark has shown us yet again the importance of rigorous
integrity in science, for example through meticulously listing and checking every conceivable
alternative explanation for an apparently positive experimental signal.
Feynman refers to this in his famous lecture on
``Cargo Cult Science" \cite{Fey},
in which he discusses what can go wrong in scientific efforts that are not
careful to maintain an appropriate level of skepticism:

\vskip 2mm

\noindent
{\it ``...there is one feature I notice that is generally missing in cargo cult science..."}

\noindent
{\it
``It's a kind of scientific integrity, a principle of scientific thought that corresponds
to a kind of utter honesty -- a kind of leaning over backwards.
For example, if you're doing an experiment, you should report everything that you
think might make it invalid -- not only what you think is right about it:
other causes that could possibly explain your results; and things you thought of
that you've eliminated by some other experiment, and how they worked --
to make sure the other fellow can tell they have been eliminated."}

\vskip 2mm

The theorists' analog of this scientific integrity presumably includes being aware
of and acknowledging all relevant experimental results, especially those that may not
support the theoretical ideas being developed. This brings me to my Feynman story. More accurately
it is an anti-Feynman story, since it ends with him stating the opposite of his principle of
scientific ethics for effect.

As a graduate student in Caltech in the mid-1970s, I~and several of the other theory students
supported ourselves as TAs by grading graduate-level physics classes. One that I graded while
it was taught by my advisor Jon Mathews was PHY205, which was an advanced quantum mechanics
class. During a two week period in which Mathews was not available he asked
Feynman to substitute for him. Of course we graduate students were excited about Feynman
teaching the class, although in some ways he was not the most widely appreciated graduate
physics instructor at Caltech; his notes were typically rather incomplete, so his in-class
derivations would have mistakes that he would correct by fudging signs and changing overall factors,
based on what he knew the right answer should be at the end. This made for an interesting lecture,
but taking notes was rather difficult. Feynman decided to teach us about
the quark model in this class, including meson and baryon state vectors and some resulting predictions.
In the final stages of these lectures we were told about magnetic moments, which was
a famous success of the quark model. Indeed, Feynman stated that he first believed in the
quark model when the baryon magnetic moments started coming out right.

This topic was especially exciting for me, since I had been calculating off-diagonal magnetic
transition moments among mesons in the quark model, and I knew that there were new experimental results 
that were in clear disagreement with the quark model predictions. So, as a dutiful TA for the class, 
I spoke up and noted that there were also interesting transition moments; Feynman agreed with this, 
and made a few statements about them as well. I then stated that there were new experimental results 
that disagreed with those predictions. He was evidently not aware of this. After class I gathered up 
these new experimental papers and went to Feynman's office, thinking that he would be very interested 
in hearing about these results. I knocked on the door, 
Feynman opened it, gave me a mock-angry smile, and said

\begin{center}
{\it ``What's the idea, messing up a perfectly good lecture by referring to experiment?"}
\end{center}

\section*{Acknowledgments}

I would like to thank the organisers for their kind invitation to
give this Meson2006 (theory) summary talk, and for the opportunity to discuss
the physics of hadrons with my fellow participants.
This research was supported in part by the
U.S. National Science Foundation through grant NSF-PHY-0244786 at the
University of Tennessee, and the U.S. Department of Energy under contract
DE-AC05-00OR22725 at Oak Ridge National Laboratory.

\end{document}